\documentclass[useAMS,usenatbib,usegraphicx]{mn2e}
\usepackage{aasmacros}

\title[Satellite Galaxies in WDM]{The haloes of bright satellite
galaxies in a warm dark matter universe}
\author[M. R. Lovell et al]{Mark R. Lovell$^{1}$\thanks{E-mail: m.r.lovell@durham.ac.uk}, Vincent Eke$^{1}$, Carlos S. Frenk$^{1}$,
Liang Gao$^{2,1}$, Adrian Jenkins$^{1}$, \newauthor Tom
Theuns$^{1,3}$, Jie Wang$^{1}$, Simon D. M. White$^{4}$, Alexey Boyarsky$^{5,6}$, \newauthor and Oleg Ruchayskiy$^{7}$\\ 
$^{1}$Institute for Computational Cosmology, Durham University, South
Road, Durham, UK, DH1 3LE\\ 
$^{2}$National Astronomical Observatories, Chinese Academy of Science, Beijing, 100012, China\\
$^{3}$Department of Physics, University of Antwerp, Groenenborgerlaan 171, B-2020 Antwerpen, Belgium\\
$^{4}$Max-Planck-Institut f\"ur Astrophysik, Karl-Schwarzschild-Stra\ss e 1, 85740 Garching bei M\"unchen, Germany\\
$^{5}$\'Ecole Polytechnique F\'ed\'erale de Lausanne, FSB/ITP/LPPC, BSP 720, CH-1015,Lausanne, Switzerland\\
$^{6}$Bogolyubov Institute of Theoretical Physics, Kiev 03680, Ukraine\\
$^{7}$\ CERN Physics Department, Theory Division, CH-1211 Geneva 23, Switzerland}

\newcommand{\Msun}{\mathrm{M_{\sun}}}
\newcommand{\lcdm}{$\Lambda$CDM}
\newcommand{\kms}{$\rmn{km\,s^{-1}}$} 
\def\gsim{ \lower .75ex \hbox{$\sim$} \llap{\raise .27ex \hbox{$>$}} }
\def\lsim{ \lower .75ex \hbox{$\sim$} \llap{\raise .27ex \hbox{$<$}} }

\begin{document}

\date{Accepted 2011 November 13. Received 2011 October 21; in original
  form 2011 April 15} 

\pagerange{\pageref{firstpage}--\pageref{lastpage}} \pubyear{2011}

\maketitle

\label{first page}

\begin{abstract}
High resolution N-body simulations of galactic cold dark matter haloes
indicate that we should expect to find a few satellite galaxies around
the Milky Way whose haloes have a maximum circular velocity in excess
of 40~\kms. Yet, with the exception of the Magellanic Clouds and the
Sagittarius dwarf, which likely reside in subhaloes with significantly
larger velocities than this, the bright satellites of the Milky Way
all appear to reside in subhaloes with maximum circular velocities
below 40~\kms. As recently highlighted by Boylan-Kolchin et al., this
discrepancy implies that the majority of the most massive subhaloes
within a cold dark matter galactic halo are too concentrated to
be consistent with the kinematic data for the bright Milky Way
satellites. Here we show that no such discrepancy exists if haloes are
made of warm, rather than cold dark matter because these haloes are
less concentrated on account of their typically later formation
epochs. Warm dark matter is one of several possible explanations for
the observed kinematics of the satellites.

\end{abstract}


\section{Introduction}

Measurements of temperature anisotropies in the microwave background
radiation \citep[e.g.][]{wmap11}, of galaxy clustering on large scales
\citep[e.g.][]{cole05}, and 
of the currently accelerated expansion of the
Universe \citep[e.g][]{Clocchiatti2006,Guy2010} have confirmed the
``Lambda cold dark matter''
($\Lambda$CDM) model, first explored theoretically 25 years ago
\citep{defw85}, as the standard model of cosmogony. These
observations probe a large range of scales, from $\sim 1$Gpc to $\sim
10$Mpc. On smaller scales, where the distribution of dark matter is
strongly non linear, observational tests of the model are more
complicated because of the complexity added by
galaxy formation processes. However, it is precisely on these scales
that the nature of the dark matter may be most clearly manifest. For
example, if the dark matter is made of warm, rather than cold
particles, free streaming in the early universe would have erased
primordial fluctuations below a scale that depends on the mass of the
dark matter particle but could be of order $10^9-10^{10}\Msun$. These
mass scales correspond to dwarf galaxies and so, in principle, the
abundance and properties of dwarf galaxies could encode information
about the nature of the dark matter.

The validity of the \lcdm\ model on galactic and subgalactic scales
has been a subject of debate for many years. Initially \cite{klypin99}
and \cite{moore99} pointed out a large discrepancy between the number
of dark matter substructures, or subhaloes, that survive inside a
galactic halo and the number of satellites that are observed around
the Milky Way. This so-called `satellite problem' is often
interpreted as indicating that the model requires most of the subhaloes
to contain no visible satellite. This aspect of the problem,
however, is readily solved by invoking the known physics of galaxy
formation, particularly early reionization of the intergalactic medium
and supernovae feedback, which inevitably inhibit the formation of
stars in small mass haloes. Detailed models that reconcile theory and
observations in this way date back to the early 2000s
\citep{Bullock_00,Benson02_sats,Somerville02}.

The paucity of observed bright satellites, however, is only one aspect
of the satellite problem. As already emphasized by 
\cite{klypin99} and \cite{moore99}, there is a problem not only with
the abundance of satellites, but also with their distribution of
circular velocities. In a halo like that of the Milky Way, the 
\lcdm\ model predicts the existence of several 
subhaloes with maximum circular velocities, $V_{\mathrm {
max}}$\footnote{ The circular velocity is given by $V=\mathrm({G}M(< r)/r)^{1/2}$,
where $M$ is the mass enclosed within radius $r$ and $\mathrm{G}$ is the
universal gravitational constant; the value of $r$ at which the
maximum of this function, $V_{\mathrm{max}}$, occurs is denoted by
$r_{\mathrm{max}}$}, in excess of $\sim 40$\kms. Using the
high-resolution simulations of galactic haloes of the Aquarius project
\citep{sp08b}, \cite{Strigari10} have recently demonstrated that it is possible to
find \lcdm\ subhaloes that accurately match the observed stellar
kinematics of the five well-studied satellites of the Milky Way. The
best fits, however, invariably have $V_{\mathrm{max}}\lsim 40$\kms. [The
Strigari et al. sample excludes the Large and Small Magellanic Clouds (LMC and SMC) which reside in more
massive haloes, and Sagittarius which is currently being disrupted.]

The discrepancy between the predicted and inferred distributions of
$V_{\mathrm{max}}$ values has recently been highlighted by
\cite{BoylanKolchin11}. Using also the Aquarius haloes, as well as the Via Lactea
simulations \citep{Madau08}, they show explicitly that the simulated
haloes typically contain a few subhaloes which are too massive and too
dense (as indicated by their value of
$V_{\mathrm{max}}/r_{\mathrm{max}}$) to host any of the observed
satellites. If such objects existed in the Milky Way, they would have
to be empty of stars despite their mass. This seems very unlikely so,
unless the Milky Way is atypical, 
there is an apparent discrepancy between model and observations. 

That the Milky Way is not typical of isolated galaxies of similar
luminosity and colour has recently been established from SDSS
data. \cite{Liu11} have shown that only 3.5 per cent of such galaxies have 2
satellites as bright as the Magellanic Clouds, while \cite{Guo11} have
shown that the luminosity function of the bright ($M_V<-14$) Milky Way
satellites has about twice the amplitude of the mean for similar
galaxies \citep[see also][]{Lares11}. While these measurements show
that the Milky Way is not an average galaxy, it is not at present
possible to compare the distribution of $V_{\mathrm {max}}$ of its
satellites with that of similar galaxies directly. However, an
indirect probe of this distribution can be constructed by combining
N-body simulations with a subhalo abundance matching procedure
\citep{Busha11}. 

In this paper we explore whether an alternative hypothesis for the
nature of the dark matter, a warm rather than a cold particle, can
provide a better match to the inferred distribution of satellite
circular velocities or masses. Specifically, we test a model in which
the dark matter is one of the particles predicted by the `neutrino
minimal standard model ($\nu$MSM)' of \citet{Asaka_05} and
\citet{by09}. In this model there is a triplet of sterile neutrinos,
the lightest of which could become nonrelativistic at a redshift of
$\sim10^{6}$, have a mass of $\sim2\mathrm{keV}$, and behave as warm
dark matter (WDM). This model is consistent with astrophysical and
particle physics data, including constraints on neutrino masses from
the Lyman-$\alpha$ forest \citep{Boyarsky09}. 

To investigate this WDM model we have resimulated one of the Aquarius
N-body haloes (Aq-A) with the power spectrum suppressed at small
scales, as expected in the WDM case. N-body simulations of
galactic and cluster WDM haloes were first carried out in the early
2000s \citep{Colin00,bode01,Knebe02}. These studies found that fewer subhaloes
form than in the CDM case and that these tend to be less concentrated
than their CDM counterparts. Qualitatively, we find similar results
but the conclusions of these early simulations are difficult to
interpret because, as we shall see later, the sharp cutoff in the
power spectrum gives rise to the formation of a large number of
artificial haloes that are purely numerical in origin
\citep{ww07}. More recently,
\cite{Maccio10} carried out new simulations of WDM models and
found that the luminosity function of satellites can be reproduced in
these models just as well as it can in the CDM case.

Our simulations have orders of magnitude higher resolution than
previous ones, enough to investigate reliably the inner structure of
the galactic subhaloes that are potential hosts of the dwarf
satellites. Furthermore, we carry out convergence tests of our results
and develop a method for distinguishing genuine WDM haloes from the
spurious objects that inevitably form in simulations of this kind. We
describe our simulations in Section~\ref{Sim}, present our results in
Section~\ref{Res}, and conclude in Section~\ref{Con}.

\section{The simulations}
\label{Sim}

\begin{table}
  \begin{tabular}{ccccr} \hline Name & $m_{\mathrm{p}}$ [M$_{\odot}$] &
  $r_{200}$ [kpc] & $M_{200}$ [M$_{\odot}$] & $N_{\mathrm{s}}$ \\ \hline Aq-A2
  & $1.370\times10^{4}$ & 245.88 & $1.842\times10^{12}$ & 30177 \\
  Aq-A3 & $4.911\times10^{4}$ & 245.64 & $1.836\times10^{12}$ & 9489
  \\ \hline Aq-AW2 & $1.370\times10^{4}$ & 242.87 &
  $1.775\times10^{12}$ & 689 \\ Aq-AW3 & $4.911\times10^{4}$ & 242.98
  & $1.778\times10^{12}$ & 338 \\ Aq-AW4 & $3.929\times10^{5}$ & 242.90
  & $1.776\times10^{12}$ & 126\\\hline \end{tabular} 
  \caption{Basic
  parameters of the simulations analysed in this paper.  The top two
  simulations are taken from the Aquarius sample of CDM dark matter
  haloes published in \citet{sp08b}. The simulations are of a single
  halo, Aq-A, at different numerical resolutions. The bottom three are
  WDM counterparts to the CDM simulations, as described
  in the main text.  The second to fifth columns give the particle
  mass ($m_{\mathrm{p}}$), the radius of the sphere of density 200 times the
  critical density ($r_{200}$), the halo mass within $r_{200}$
  ($M_{200}$) and the number of subhaloes within the main halo
  ($N_{\mathrm{s}}$). The smallest subhaloes, determined by \textsc{subfind}, contain 20
  particles.}

  \label{table1}
\end{table}

To compare the properties of subhaloes in Milky Way mass haloes in CDM
and WDM universes, we have assembled a sample of five high resolution
simulations of galactic mass haloes. All the simulations have the same
basic cosmological parameters: in units of the critical density, a
total matter density, $\Omega_{m}=0.25$ and a cosmological constant,
$\Omega_{\Lambda}=0.75$.  The linear power spectrum has a spectral
index $n_s=1$ and is normalised to
give $\sigma_{8}=0.9$, with
$H_{0}=100h\mathrm{kms^{-1}Mpc^{-1}}=73\mathrm{kms^{-1}Mpc^{-1}}$
\citep{sp08b}. \footnote{Although this
set of parameters is discrepant at about the 3$\sigma$ level with the
latest constraints from microwave background and large-scale structure
data \citep{wmap11}, particularly with the values of $\sigma_{8}$ and
$n_s$, the differences are not important for our purposes. For
example, \cite{BoylanKolchin11} show that the structure of Aquarius
subhaloes is statistically similar to that of subhaloes in the Via
Lactea simulations which assume a value of $\sigma_8=0.74$, lower 
than that of \cite{wmap11}, and a spectral index of 0.95.}

We have taken two simulations from the Aquarius project described in
\cite{sp08b}, both of the same halo, Aq-A, but of different
resolution, corresponding to levels~2 and~3 in the notation of
\cite{sp08b}.  The higher resolution, level~2, simulation has more
than a hundred million particles within $r_{200}$, the radius of a
sphere about the halo centre, encompassing a mean density of 200 times
the critical density.  The level~3 simulation has 3.6 times fewer
particles. In both cases, the mass of the halo within $r_{200}$ is
about $1.8\times10^{12}M_\odot$, which is consistent with the
estimated mass of the Milky Way \citep{Li_08, Xue_08, Gnedin_10}. The
basic properties of these haloes are given at the top of
Table~\ref{table1}. Substructures were identified using the \textsc{subfind}
algorithm \citep{Springel01} to find gravitationally bound subhaloes
within them.

\begin{figure}
   \includegraphics[angle=-90, scale=0.33]{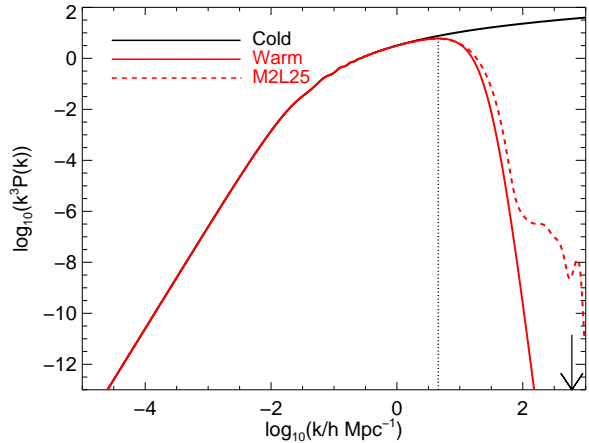}
  \caption{The solid lines show the linear power spectra \citep[from
\textsc{cmbfast};][]{Sel_Zal96} used for the two simulations. Black is
the original, CDM Aq-A spectrum, and red is that of Aq-AW. The
vertical dashed line marks the peak of the WDM spectrum peak. The
arrow marks the Nyquist frequency of the level 2 simulations. The
dashed red curve corresponds to the M2L25 model of \citep{Boyarsky09}
which is almost identical to the solid red curve for scales below
$k\sim 10$~$h\mathrm{Mpc^{-1}}$. }
  \label{Pwsp}
\end{figure}

We created three WDM counterparts to the CDM haloes by
running new simulations using the same code and numerical parameters
as \cite{sp08b} but with WDM initial conditions.  The WDM initial
conditions were created keeping the same phases and the same
unperturbed particle positions as in the CDM case, but using a WDM
matter power spectrum instead to scale the amplitudes of the
fluctuations.  The linear matter power spectrum for both the CDM and
WDM simulations is shown in Fig.~\ref{Pwsp} with solid lines adopting
an arbitrary normalisation at large scales.

\begin{figure}
   \includegraphics[angle=0, scale=0.4]{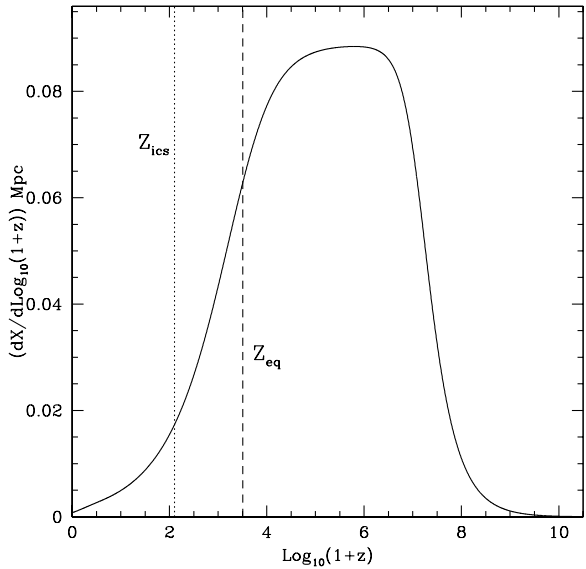}
  \caption{The free streaming comoving distance traveled per log
interval of $1+z$, where $z$ is redshift, for a WDM
particle with a fiducial velocity of $24~{\rm m}~{\rm s}^{-1}$ at the
present day.  The dashed vertical line marks the redshift of
matter-radiation equality. The dotted vertical line indicates the
start redshift of the WDM simulations. }
  \label{Fstream}
\end{figure}

The WDM power spectrum has a strong cut off at high
wavenumbers due to the free streaming of the WDM
particles.  In an unperturbed universe at the present day the typical
velocities of WDM particles are only a few tens of
$\mathrm{ms^{-1}}$. This implies that the particles ceased to be relativistic
after a redshift of $z\sim10^7$, well before the end of the
radiation-dominated 
era, as suggested by the word `warm'.  Fig.~\ref{Fstream} illustrates
the free streaming of a typical WDM particle over cosmic
time.  The area under the curve is the comoving distance traveled.  It
is evident that the WDM particle travels the greatest
comoving distance during the radiation-dominated era after it has become
nonrelativistic \citep{bode01}.  Over the duration of the N-body
simulation, which starts at $z=127$, a particle typically travels a
distance of around $\mathrm{14~kpc}$, which is small compared to the
total distance from early times of $\mathrm{400~kpc}$.  For
comparison, the mean interparticle separation for the high resolution
region in our highest resolution simulation is $\mathrm{7.4~kpc}$,
similar to the free-streaming distance traveled by the particles after
$z= 127$. This means that the effects of streaming during the
simulation are small, and only affect scales that are barely resolved
in our simulations. For this reason we chose to set the particle
velocities in the same way as in the CDM case, where the particle
velocity is a function of the unperturbed comoving position of a
particle and is determined solely by the matter fluctuations.

 The WDM matter power spectrum we assume has a shape
characteristic of a `thermal relic' (Bode et al. 2001). However our
WDM matter power spectrum is also an excellent fit for scales below
$k\sim 10$~$h\mathrm{Mpc^{-1}}$, to the matter power spectrum of the M2L25 model of
\cite{Boyarsky09}, which is shown as a dashed line in Fig.~\ref{Pwsp}.
At $k = 10$~$h`mathrm{Mpc^{-1}}$ the power in both WDM curves is a factor three below
that of CDM and falls away very rapidly beyond here in both models.
The M2L25 model corresponds to a \emph{resonantly produced} ∼ 2keV
sterile neutrino with a highly non-equilibrium spectrum of primordial
velocities.  The model is only just consistent with astrophysical
constraints \citep{Boyarsky09} and so maximizes the differences
between the substructures in the CDM and WDM haloes,
both in their internal structure and in their abundance.

 For wavenumbers below the peak at  $4.5 h{\rm Mpc}^{-1}$ the linear
 WDM power spectrum is well approximated by the product of the linear
CDM power spectrum times the square of the Fourier
transform of a spherical top-hat filter of unit amplitude and 
radius $\mathrm{320~kpc}$, or equivalently, containing a
mass of $5\times10^{9}\mathrm{M}_{\odot}$ at the mean density.

\begin{figure*}
\includegraphics[scale=0.90]{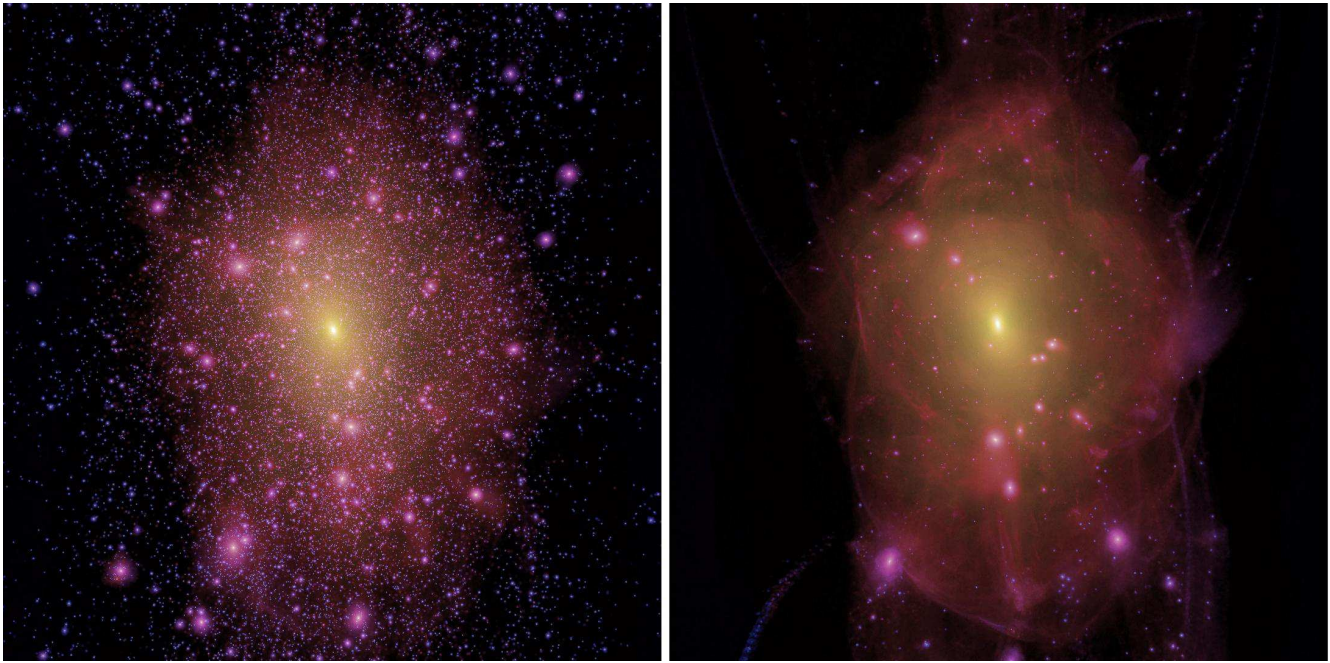}
\caption{ Images of the CDM (left) and WDM (right) level~2 haloes at
$z=0$. Intensity indicates the line-of-sight projected square of the
    density, and hue the projected density-weighted velocity dispersion,
ranging from blue (low velocity dispersion) to yellow (high velocity
dispersion). Each box is 1.5 Mpc on a side.  Note
the sharp caustics visible at large radii in the WDM image, several of
which are also present, although less well defined, in the CDM case. }
\label{DensityVelDisp}
\end{figure*}

Images of the CDM and WDM haloes are shown in
Fig.~\ref{DensityVelDisp}. 
As shown in Table~\ref{table1}, the mass of the main halo in the WDM
simulation is very similar to that of the CDM halo, just a few per cent
lighter.  However, the number of substructures in the WDM case is much
lower, reflecting the fact that the small scale power in these
simulations is greatly reduced. Some of the largest subhaloes can be
matched by eye in the images of the two simulations. 

\cite{sp08b} showed that it is possible to make precise matches
between substructures at different resolutions for the Aq-A halo,
allowing the numerical convergence of properties of substructures to
be checked for individual substructures.  For this paper, we have
found matches between subhaloes in the Aq-AW2, Aq-AW3, and
Aq-AW4 simulations. We make these matches at the epoch when
the subhaloes first have a mass which is more than half the
mass they have at the time when they first infall into the main halo
(which is very close to the maximum mass they ever attain). At this
epoch it is relatively easy to match the largest substructures in
these three simulations as the corresponding objects have very similar
positions, velocities and masses.

The number of subhaloes that can be matched between the two WDM
simulations is much smaller than that between the corresponding CDM
simulations, and is also a much smaller fraction of the total number
of subhaloes identified by \textsc{subfind}.  The majority of
substructures identified in the WDM simulations form through
fragmentation of the sharply delineated filaments characteristic of
WDM simulations and do not have counterparts in the simulations of
different resolution. The same phenomenon is seen in hot dark
matter simulations and is numerical in origin, occurring along the
filaments on a scale matching the interparticle separation
\citep{ww07}. This artificial fragmentation is apparent in
Fig.~\ref{DensityVelDisp}.

 We will present a detailed description of subhalo matching in
a subsequent paper but, in essence, we have found that matching
subhaloes works best when comparing the Lagrangian regions of the
initial conditions from which the subhaloes form, rather than the
subhaloes themselves. We use a sample of the particles present in a
subhalo at the epoch when it had half of the mass at infall to define
the Lagrangian region from which it formed.  We have devised a
quantitative measure of how well the Lagrangian regions of the
substructures overlap between the simulations of different resolution,
and select as genuine only those subhaloes with strong matches between
all three resolutions. We find that these criteria identify a sample
of fifteen relatively massive subhaloes with mass at infall greater
than $2\times10^{9}M_{\odot}$, together with a few more subhaloes with
infall mass below $10^{9}\mathrm{M}_{\odot}$.  This sample of fifteen
subhaloes includes all of the subhaloes with infall masses greater
$10^{9}\mathrm{M}_{\odot}$.

 We have also found that the shapes of the Lagrangian regions of
 spurious haloes in our WDM simulations are typically very aspherical. We have therefore devised a second measure
 based on sphericity as an independent way to reject spurious haloes.
 All fifteen of the massive subhaloes identified by the first criterion
 pass our shape test, but all but one subhalo with an infall mass below
 $10^{9}\mathrm{M}_{\odot}$ are excluded.  For the purposes of this
 paper we need only the 12 most massive subhaloes at infall to make
 comparisons with the Milky Way satellites. 

 For both our WDM and CDM catalogues, we select a sample
made up of  the 12 most massive subhaloes at infall found today within
$\mathrm{300~kpc}$ of the main halo centre.  In the Aq-AW2 simulation
these subhaloes are resolved with between about 2 and 0.23
million particles at their maximum mass. We use the particle nearest
the centre of the gravitational potential to define the centre of each
subhalo and hence determine the values of $V_{\mathrm{max}}$ and
$r_{\mathrm{max}}$ defined in Section~1.

\section{Results}
\label{Res}

In this section, we study the central masses of the substructures
found within 300 kpc of the centres of the CDM and WDM Milky Way like
haloes. These results are compared with the masses within the
half-light radii, inferred by \citet{Walker09, Walker10} and \citet{wo10} from kinematic
measurements, for the 9 bright ($L_V>10^5 L_\odot$) Milky Way dwarf
spheroidal galaxies.

\begin{figure}
   \includegraphics[angle=-90, scale=0.33]{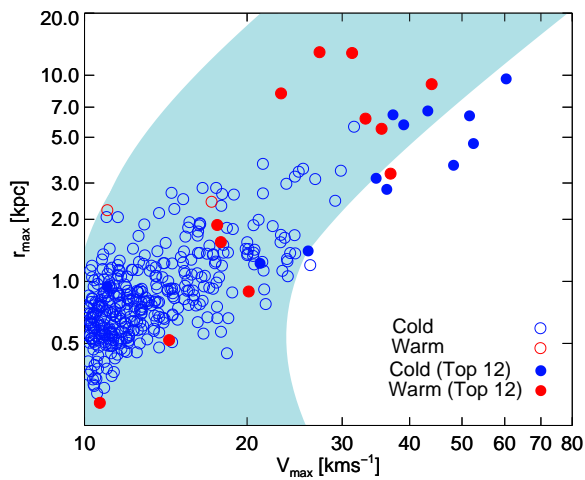}
   \caption{The correlation between subhalo maximum circular velocity
   and the radius at which this maximum occurs. Subhaloes lying within
   300kpc of the main halo centre are included. The 12 CDM and WDM subhaloes with the most massive progenitors are shown as blue and red filled circles respectively; the remaining subhaloes are shown as empty circles. The shaded area represents the $2\sigma$
   confidence region for possible hosts of the 9 bright Milky Way
   dwarf spheroidals determined by \citet{BoylanKolchin11}.}  
\label{VvsR}
\end{figure}

Following the study by \citet{BoylanKolchin11}, in Fig.~\ref{VvsR} we plot the
correlation between $V_{\mathrm{max}}$ and $r_{\mathrm{max}}$ for the
subhaloes in Aq-AW2 and Aq-A2 that lie within 300kpc of the centre of
the main halo. Only those WDM subhaloes selected using our matching scheme
are included, whereas all Aq-A2 subhaloes are
shown. The CDM subhaloes are a subset of those shown in fig.~2 of
\citet{BoylanKolchin11}, and show $V_{\mathrm{max}}$ values that are typically
$\sim 50$ per cent larger than those of WDM haloes with a similar
$r_{\mathrm{max}}$. By assuming that the mass density in the subhaloes
containing the observed dwarf spheroidals follows an NFW profile
\citep{NFW_96,NFW_97}, \citet{BoylanKolchin11} found the locus of possible
$(r_{\mathrm{max}},V_{\mathrm{max}})$ pairs that are consistent with
the observed half-light radii and their enclosed masses.  This is
represented by the shaded region in Fig.~\ref{VvsR}. As
\citet{BoylanKolchin11} observed with their larger sample, several of the largest CDM
subhaloes have higher maximum circular velocities than appears to be
the case for the Milky Way bright dwarf spheroidals. By contrast, the
largest WDM subhaloes are consistent with the Milky Way data.

Rather than assuming a functional form for the mass density
profile in the observed subhaloes, a more direct approach is to
compare the observed masses within the half-light radii of the dwarf
spheroidals with the masses within the same radii in the simulated
subhaloes. To provide a fair comparison we must choose the simulated
subhaloes that are most likely to correspond to those that host the 9
bright dwarf spheroidals in the Milky Way. As stripping of subhaloes
preferentially removes dark matter relative to the more centrally
concentrated stellar component, we choose to associate final satellite
luminosity with the maximum progenitor mass for each surviving
subhalo. This is essentially the mass of the object as it falls into
the main halo. The smallest subhalo in each of our samples has
an infall mass of $3.2\times10^{9}\mathrm{M_{\odot}}$ in the WDM case, and
$6.0\times10^{9}\mathrm{M_{\odot}}$ in the CDM case. 

\begin{figure}
  \includegraphics[angle=-90, scale=0.33]{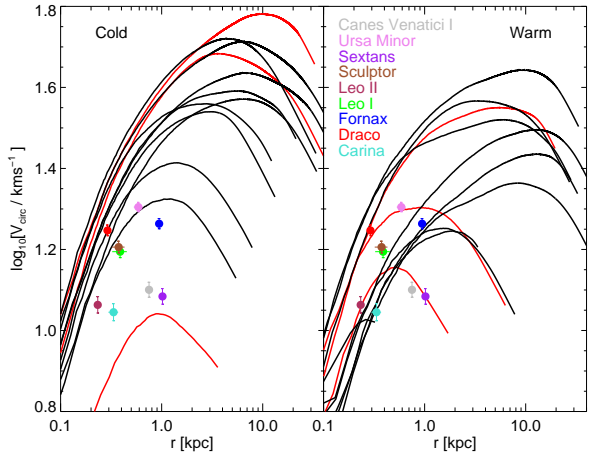}
  \caption{Circular velocity curves for the 12 CDM (left)
  and WDM (right) subhaloes that had the most massive
  progenitors. The 3 red curves represent subhaloes with the most
  massive progenitors, which could correspond to those currently
  hosting counterparts of the LMC, SMC and
  the Sagittarius dwarf. The 9 black curves might more fairly be
  compared with the data for the 9 bright dwarf spheroidal galaxies of
  the Milky Way considered by \citet{wo10}. Deprojected half-light
  radii and their corresponding half-light masses, as determined by
  \citet{wo10} from line-of-sight velocity measurements, are
used to derive the half-light circular velocities of each
dwarf spheroidal. These velocities and radii are shown as  coloured
points. The legend indicates the colour coding of the
different galaxies.}
    \label{MVB}
\end{figure}

The LMC, SMC and the Sagittarius dwarf are
all more luminous than the 9 dwarf spheroidals considered by
\citet{BoylanKolchin11} and by us. As noted above, the Milky Way is exceptional
in hosting galaxies as bright as the Magellanic Clouds, while
Sagittarius is in the process of being disrupted so its current mass
is difficult to estimate. Boylan-Kolchin et al. hypothesize that these
three galaxies all have values of
$V_{\mathrm{max}}>60\mathrm{kms^{-1}}$ at infall and exclude simulated
subhaloes that have these values at infall as well as
$V_{\mathrm{max}}>40\mathrm{kms^{-1}}$ at the present day from their
analysis. In what follows, we retain all subhaloes but, where
appropriate, we highlight those that might host large satellites akin
to the Magellanic Clouds and Sagittarius. 

The circular velocity curves at $z=0$ for the 12 subhaloes which
had the most massive progenitors at infall are shown in Fig.~\ref{MVB}
for both WDM and CDM. The circular velocities within the
half-light radius of the 9 satellites measured by \cite{wo10}
are also plotted as symbols. Leo-II has the smallest half-light radius,
$\sim 200$pc. To compare the satellite data with the simulations we
must first check the convergence of the simulated subhalo masses
within at least this radius. We find that the median of the ratio of
the mass within 200pc in the Aq-W2 and Aq-W3 simulations is
$W2/W3\sim 1.22$, i.e., the mass within 200pc in the Aq-W2
simulation has converged to better than
$\sim 22\%$.

As can be inferred from Fig.~\ref{MVB}, the WDM subhaloes have
similar central masses to the observed satellite galaxies, while the
CDM subhaloes are almost all too massive at the corresponding
radii. The CDM subhaloes have central masses that are typically 2-3
times larger than the Milky Way satellites. There is one CDM subhalo
that lies at lower masses than all 9 dwarf spheroidals, but this had
one of the three most massive progenitors and has been almost
completely destroyed by tidal forces.

\begin{figure}
  \includegraphics[angle=-90, scale=0.33]{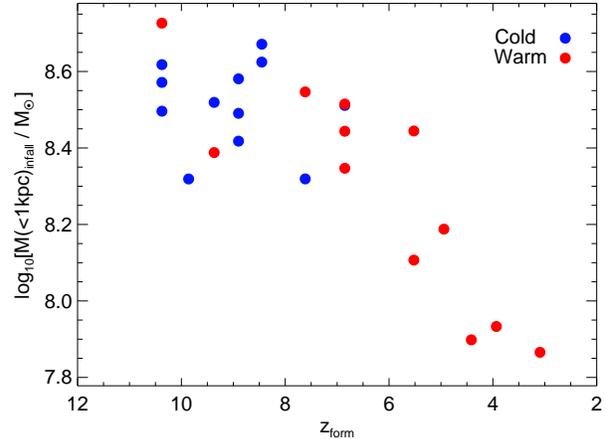}
  \caption{The correlation between subhalo central mass at infall and
  the redshift of formation, $z_{\rm form}$, defined as the redshift at
  which the total mass of each proto subhalo first exceeded this 
  value. Central mass is defined within 1 kpc, and CDM and WDM results
  are shown with blue and red symbols respectively.}  
\label{zvM}
\end{figure}

Fig.~\ref{VvsR} and~\ref{MVB} show that the WDM subhaloes are
less centrally concentrated than those in the corresponding CDM
halo. Concentrations typically reflect the epoch at which the halo
formed \citep{NFW_96,NFW_97,ek01}. To investigate systematic
differences in the formation epoch of the WDM and CDM subhaloes in our
sample, we must choose a suitable definition of formation time. Since
we are considering only the central mass, and we do not wish to
introduce scatter in any correlation by using subhaloes that may have
been stripped, we define the formation time as the first time at which
the total progenitor mass exceeds the mass within 1 kpc at infall. The
correlation of this redshift with the mass within 1 kpc at infall is
shown in Fig.~\ref{zvM} for the 12 most massive WDM and CDM
progenitors that survive to $z=0$ as distinct subhaloes. Evidently, the
proto subhaloes that form later, which are generally WDM not CDM ones,
have the lowest central masses. The mean difference between the top 12
WDM and CDM proto-subhalo masses within 1 kpc is approximately a
factor 2.

\begin{figure}
  \includegraphics[angle = -90, scale=0.33]{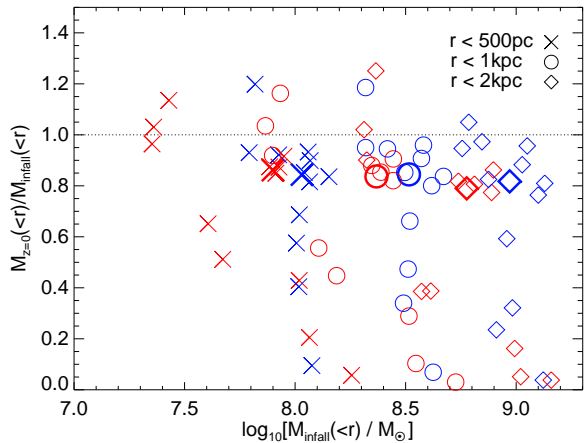}
  \caption{The variation with subhalo mass at infall of the
  ratio of the present day mass to the infall mass contained within
  500pc, 1kpc and 2kpc. Data are shown for the 12 subhaloes identified
  at $z=0$ which had the most massive progenitors, with CDM in blue
  and WDM in red. The symbol type denotes the radius interior to which
  the central mass is being measured and large symbols show the
  medians of the corresponding distributions. We find no
systematic differences between the CDM and WDM subhalo mass ratios.}

\label{MVM}
\end{figure}

Because of their later formation time, the infalling WDM subhaloes
already have lower central masses than those falling into the
corresponding CDM haloes. As their mass is less centrally
concentrated, the WDM subhaloes are more susceptible to
stripping. While this is most important in the outer regions of the
subhaloes, the mass profiles in Fig.~\ref{MVB} show that the inner
regions of some of the subhaloes have also endured significant
depletion since infall. Fig.~\ref{MVM} shows, for both WDM and CDM
subhaloes, the ratio, $M_{z=0}(<r)/M_{\rm infall}$, of the present
day mass contained within $r=0.5$, 1 and 2 kpc to the mass at infall,
as a function of the central mass at infall at the chosen radius.
 On average, the median mass at infall for WDM is lower by
$\sim0.15$ dex than the corresponding mass for CDM. One subhalo gains mass
between infall and $z=0$ because it accretes another subhalo.  While
there is a large scatter among the different subhaloes, with some
having lost the majority of their central mass since infall, no
significant systematic difference between WDM and CDM subhaloes is
apparent.  This implies that the reason why the WDM subhaloes provide
a better fit to the half-light masses of the 9 Milky Way dwarf
spheroidals studied by \citet{wo10} is not excess stripping
but the later formation time, and correspondingly typical lower
concentration, of the WDM proto subhaloes compared to their
CDM counterparts.

\section{Discussion and conclusions}
\label{Con}

The properties of the satellite galaxies of the Milky Way have posed a
longstanding puzzle for CDM theories of galaxy formation.
Two aspects of this puzzle have reportedly been separately and
independently solved. One is the luminosity function of the
satellites. The basic idea - the suppression of galaxy formation in
small haloes by a combination of feedback effects produced by the
reionization of gas at high redshift and supernova heating - was
suggested by \cite{Kauffmann1993} and explored thoroughly in the early
2000s \citep{Bullock_00,Benson02_sats,Somerville02} and has been
revisited many times since then \citep[see][and references therein for
the most recent discussion]{Font11}. The other aspect concerns the
dynamical state of the satellites. \cite{Strigari10} have shown that
there exist subhaloes in the Aquarius CDM simulations
that fit the stellar spectroscopic data for the well-studied
satellites extremely well.

There is a third aspect to the puzzle, however, that has not yet been
fully addressed and this is whether the CDM models that
account for the satellite luminosity function also account for the
satellites' internal dynamics. In other words, do the models assign
the correct luminosities to subhaloes with the correct dynamics? At
face value, the answer seems to be `no'. This is already evident in
the analysis of \cite{Strigari10} in which the best fit dynamical
models imply velocity dispersions (or equivalently $V_{\mathrm{max}}$
values) for the brightest dwarf spheroidals that are smaller than the
velocity dispersions of the largest subhaloes. It is this discrepancy
that has recently been highlighted by \cite{BoylanKolchin11}.

In this paper, we have compared a high resolution N-body simulation of
one of the Aquarius galactic haloes with a WDM
counterpart. The initial conditions for both had the same phases and
the same unperturbed particle positions. For the WDM
simulation we chose a form of the power spectrum corresponding to one
of the models discussed by \citet{Asaka_05} and
\citet{by09}, in which the dark matter is a sterile neutrino with 
mass $\sim2\mathrm{keV}$, just consistent with various astrophysical
constraints \citep{Boyarsky09}. The suppression of the power spectrum
at masses below $\sim 10^{10}\Msun$\ delays the formation of the
haloes that will end up hosting the satellites and, as we have shown,
this lowers their concentration compared to that of the corresponding
CDM haloes. This is enough to reconcile the dynamics of
the subhaloes with the data.

While a WDM model naturally produces haloes that are less
concentrated than their CDM counterparts, this is only
one possible solution to the puzzle. Other forms of dark matter such
as `meta-CDM' resulting from the decay of cold thermal
relics could produce a similar outcome \citep{Strigari07}. Also, it
must be borne in mind that the values of $V_{\mathrm {max}}$ for Milky
Way satellites are not directly measured but inferred by making
assumptions about their dynamical state. If some of these assumptions
are unrealistic, this could lead to an underestimate of the values of
$V_{\mathrm {max}}$
\citep[e.g.][]{Stoehr02}. Another possibility is that the satellite
population of the Milky Way is not typical of the average to which the
model predictions apply. It has recently been shown by
\cite{Liu11}, \cite{Guo11} and \cite{Lares11} that the bright end of the
Milky Way satellite luminosity function is different from the
average. Finally, we cannot exclude the possibility that baryonic
processes occurring during the formation of satellite galaxies in the 
CDM cosmogony might have lowered the concentration of haloes, for
example, in the manner suggested by \cite{NEF96}. Recent simulations 
\citep{Read05,Mashchenko08,Governato10} suggest that these processes
could be important although it remains to be seen if they are enough
to reconcile the CDM model with the dynamics of the Milky Way
satellites.

\section*{Acknowledgements}

We thank Qi Guo and Louis Strigari for useful discussions. ML
acknowledges an STFC studentship. CSF acknowledges a Royal Society
Wolfson Research Merit Award and an ERC Advanced Investigator grant.
The simulations used in this paper were carried out on the
Supercomputer Center of the Chinese Academy of Science and the
Cosmology Machine supercomputer at the Institute for Computational
Cosmology, Durham.  LG acknowledges support from the
One hundred talents Program of the Chinese Academy of Science (CAS),
the National Basic Research Program of China (program 973, under grant
No. 2009CB24901), {\small NSFC} grant No. 10973018, and an STFC
Advanced Fellowship. The Cosmology Machine is part of the DiRAC
Facility jointly funded by STFC, the Large Facilities Capital Fund of
BIS, and Durham University. This work was supported in part by an STFC
rolling grant to the ICC.


\begin{thebibliography}{}

\bibitem[\protect\citeauthoryear{{Asaka} \& {Shaposhnikov}}{{Asaka} \&
  {Shaposhnikov}}{2005}]{Asaka_05}
{Asaka} T.,  {Shaposhnikov} M.,  2005, Physics Letters B, 620, 17

\bibitem[\protect\citeauthoryear{Benson, Frenk, Lacey, Baugh \& Cole}{Benson
  et~al.}{2002}]{Benson02_sats}
Benson A.~J.,  Frenk C.~S.,  Lacey C.~G.,  Baugh C.~M.,    Cole S.,  2002,
  MNRAS, 333, 177

\bibitem[\protect\citeauthoryear{{Bode}, {Ostriker} \& {Turok}}{{Bode}
  et~al.}{2001}]{bode01}
{Bode} P.,  {Ostriker} J.~P.,    {Turok} N.,  2001, \apj, 556, 93

\bibitem[\protect\citeauthoryear{{Boyarsky}, {Lesgourgues}, {Ruchayskiy} \&
  {Viel}}{{Boyarsky} et~al.}{2009b}]{Boyarsky09}
{Boyarsky} A.,  {Lesgourgues} J.,  {Ruchayskiy} O.,    {Viel} M.,  2009b,
  Physical Review Letters, 102, 201304

\bibitem[\protect\citeauthoryear{{Boyarsky}, {Ruchayskiy} \&
  {Shaposhnikov}}{{Boyarsky} et~al.}{2009a}]{by09}
{Boyarsky} A.,  {Ruchayskiy} O.,    {Shaposhnikov} M.,  2009a, Annual Review of
  Nuclear and Particle Science, 59, 191

\bibitem[\protect\citeauthoryear{{Boylan-Kolchin}, {Bullock} \&
  {Kaplinghat}}{{Boylan-Kolchin} et~al.}{2011}]{BoylanKolchin11}
{Boylan-Kolchin} M.,  {Bullock} J.~S.,    {Kaplinghat} M.,  2011, \mnras, 415,
  L40

\bibitem[\protect\citeauthoryear{{Bullock}, {Kravtsov} \& {Weinberg}}{{Bullock}
  et~al.}{2000}]{Bullock_00}
{Bullock} J.~S.,  {Kravtsov} A.~V.,    {Weinberg} D.~H.,  2000, \apj, 539, 517

\bibitem[\protect\citeauthoryear{{Busha}, {Wechsler}, {Behroozi}, {Gerke},
  {Klypin} \& {Primack}}{{Busha} et~al.}{2011}]{Busha11}
{Busha} M.~T.,  {Wechsler} R.~H.,  {Behroozi} P.~S.,  {Gerke}
  B.~F.,  {Klypin}  A.~A.,    {Primack} J.~R.,  2011, ApJ, 743, 117

\bibitem[\protect\citeauthoryear{{Cole}, {Percival}, {Peacock}, {Norberg},
  {Baugh}, {Frenk}, {Baldry}, {Bland-Hawthorn} \& {and 23 others}}{{Cole}
  et~al.}{2005}]{cole05}
{Cole} S., et~al.,  2005, MNRAS, 362, 505

\bibitem[\protect\citeauthoryear{Col{\'{\i}}n, Avila-Reese, 
\& Valenzuela}{2000}]{Colin00} 
Col{\'{\i}}n P., Avila-Reese V., Valenzuela O., 2000, ApJ, 542, 622

\bibitem[\protect\citeauthoryear{Clocchiatti et 
al.}{2006}]{Clocchiatti2006} 
Clocchiatti A., et al., 2006, ApJ, 642, 1

\bibitem[\protect\citeauthoryear{{Davis}, {Efstathiou}, {Frenk} \&
  {White}}{{Davis} et~al.}{1985}]{defw85}
{Davis} M.,  {Efstathiou} G.,  {Frenk} C.~S.,    {White} S.~D.~M.,  1985, \apj,
  292, 371

\bibitem[\protect\citeauthoryear{{Eke}, {Navarro} \& {Steinmetz}}{{Eke}
  et~al.}{2001}]{ek01}
{Eke} V.~R.,  {Navarro} J.~F.,    {Steinmetz} M.,  2001, \apj, 554, 114

\bibitem[\protect\citeauthoryear{{Font}}{{Font} et~al.}{2011}]{Font11}
{Font} A.~S.,  {et al.} 2011, \mnras, 417, 1260

\bibitem[\protect\citeauthoryear{{Gnedin}, {Brown}, {Geller} \&
  {Kenyon}}{{Gnedin} et~al.}{2010}]{Gnedin_10}
{Gnedin} O.~Y.,  {Brown} W.~R.,  {Geller} M.~J.,    {Kenyon} S.~J.,  2010,
  \apjl, 720, L108

\bibitem[\protect\citeauthoryear{{Governato}}{{Governato} et~al.}{2010}]{Governato10}
{Governato} F.,et~al., 2010, \nat, 463, 203


\bibitem[\protect\citeauthoryear{{Guo}, {Cole}, {Eke} \& {Frenk}}{{Guo}
  et~al.}{2011}]{Guo11}
{Guo} Q.,  {Cole} S.,  {Eke} V.,    {Frenk} C.,  2011, \mnras, 417, 370

\bibitem[\protect\citeauthoryear{Guy et 
al.}{2010}]{Guy2010}
Guy J., et al., 2010, A\&A, 523, A7 

\bibitem[\protect\citeauthoryear{Kauffmann, White, 
\& Guiderdoni}{1993}]{Kauffmann1993} Kauffmann G., White S.~D.~M., Guiderdoni B., 1993, MNRAS, 264, 201 

\bibitem[\protect\citeauthoryear{{Klypin}, {Kravtsov}, {Valenzuela} \&
  {Prada}}{{Klypin} et~al.}{1999}]{klypin99} {Klypin} A., {Kravtsov}
  A.~V., {Valenzuela} O., {Prada} F., 1999, \apj, 522, 82


\bibitem[\protect\citeauthoryear{Knebe et al.}{2002}]{Knebe02} 
Knebe A., Devriendt J.~E.~G., Mahmood A., Silk J., 2002, MNRAS, 329, 813

\bibitem[\protect\citeauthoryear{{Komatsu}, {Smith}, {Dunkley}, {Bennett},
  {Gold}, {Hinshaw}, {Jarosik}, {Larson} \& {and 13 others}}{{Komatsu}
  et~al.}{2011}]{wmap11}
{Komatsu} E., et~al.,  2011, \apjs,
  192, 18

\bibitem[\protect\citeauthoryear{{Lares}, {Lambas} \&
  {Dom{\'{\i}}nguez}}{{Lares} et~al.}{2011}]{Lares11}
{Lares} M.,  {Lambas} D.~G.,    {Dom{\'{\i}}nguez} M.~J.,  2011, \aj, 142, 13

\bibitem[\protect\citeauthoryear{{Li} \& {White}}{{Li} \&
  {White}}{2008}]{Li_08}
{Li} Y.,  {White} S.~D.~M.,  2008, \mnras, 384, 1459

\bibitem[\protect\citeauthoryear{{Liu}, {Gerke}, {Wechsler}, {Behroozi} \&
  {Busha}}{{Liu} et~al.}{2011}]{Liu11}
{Liu} L.,  {Gerke} B.~F.,  {Wechsler} R.~H.,  {Behroozi} P.~S.,    {Busha}
  M.~T.,  2011, \apj, 733, 62

\bibitem[\protect\citeauthoryear{Macci{\`o} 
\& Fontanot}{2010}]{Maccio10} 
Macci{\`o} A.~V., Fontanot F., 2010, MNRAS, 404, L16

\bibitem[\protect\citeauthoryear{{Madau}, {Diemand} \& {Kuhlen}}{{Madau}
  et~al.}{2008}]{Madau08}
{Madau} P.,  {Diemand} J.,    {Kuhlen} M.,  2008, \apj, 679, 1260

\bibitem[\protect\citeauthoryear{{Mashchenko}, {Wadsley} \&
  {Couchman}}{{Mashchenko} et~al.}{2008}]{Mashchenko08}
{Mashchenko} S.,  {Wadsley} J.,    {Couchman} H.~M.~P.,  2008, Science, 319,
  174

\bibitem[\protect\citeauthoryear{{Moore}, {Ghigna}, {Governato}, {Lake},
  {Quinn}, {Stadel} \& {Tozzi}}{{Moore} et~al.}{1999}]{moore99}
{Moore} B.,  {Ghigna} S.,  {Governato} F.,  {Lake} G.,  {Quinn} T.,  {Stadel}
  J.,    {Tozzi} P.,  1999, \apjl, 524, L19

\bibitem[\protect\citeauthoryear{{Navarro}, {Eke} \& {Frenk}}{{Navarro}
  et~al.}{1996a}]{NEF96}
{Navarro} J.~F.,  {Eke} V.~R.,    {Frenk} C.~S.,  1996a, \mnras, 283, L72

\bibitem[\protect\citeauthoryear{{Navarro}, {Frenk} \& {White}}{{Navarro}
  et~al.}{1996b}]{NFW_96}
{Navarro} J.~F.,  {Frenk} C.~S.,    {White} S.~D.~M.,  1996b, \apj, 462, 563

\bibitem[\protect\citeauthoryear{{Navarro}, {Frenk} \& {White}}{{Navarro}
  et~al.}{1997}]{NFW_97}
{Navarro} J.~F.,  {Frenk} C.~S.,    {White} S.~D.~M.,  1997, \apj, 490, 493

\bibitem[\protect\citeauthoryear{{Read} \& {Gilmore}}{{Read} \&
  {Gilmore}}{2005}]{Read05}
{Read} J.~I.,  {Gilmore} G.,  2005, \mnras, 356, 107

\bibitem[\protect\citeauthoryear{{Seljak} \& {Zaldarriaga}}{{Seljak} \&
  {Zaldarriaga}}{1996}]{Sel_Zal96}
{Seljak} U.,  {Zaldarriaga} M.,  1996, \apj, 469, 437

\bibitem[\protect\citeauthoryear{Somerville}{Somerville}{2002}]{Somerville02}
Somerville R.~S.,  2002, \apjl, 572, L23


\bibitem[\protect\citeauthoryear{{Springel}, {Wang}, {Vogelsberger}, {Ludlow},
  {Jenkins}, {Helmi}, {Navarro}, {Frenk} \& {White}}{{Springel}
  et~al.}{2008}]{sp08b}
{Springel} V.,  {Wang} J.,  {Vogelsberger} M.,  {Ludlow} A.,  {Jenkins} A.,
  {Helmi} A.,  {Navarro} J.~F.,  {Frenk} C.~S.,    {White} S.~D.~M.,  2008,
  \mnras, 391, 1685

\bibitem[\protect\citeauthoryear{{Springel}, {White}, {Tormen} \&
  {Kauffmann}}{{Springel} et~al.}{2001}]{Springel01}
{Springel} V.,  {White} S.~D.~M.,  {Tormen} G.,    {Kauffmann} G.,  2001,
  \mnras, 328, 726

\bibitem[\protect\citeauthoryear{{Stoehr}, {White}, {Tormen} \&
  {Springel}}{{Stoehr} et~al.}{2002}]{Stoehr02}
{Stoehr} F.,  {White} S.~D.~M.,  {Tormen} G.,    {Springel} V.,  2002, \mnras,
  335, L84

\bibitem[\protect\citeauthoryear{{Strigari}, {Frenk} \& {White}}{{Strigari}
  et~al.}{2010}]{Strigari10}
{Strigari} L.~E.,  {Frenk} C.~S.,    {White} S.~D.~M.,  2010, \mnras, 408, 2364

\bibitem[\protect\citeauthoryear{{Strigari}, {Kaplinghat} \&
  {Bullock}}{{Strigari} et~al.}{2007}]{Strigari07}
{Strigari} L.~E.,  {Kaplinghat} M.,    {Bullock} J.~S.,  2007, \prd, 75, 061303

\bibitem[\protect\citeauthoryear{{Walker}, {Mateo}, {Olszewski},
  {Pe{\~n}arrubia}, {Wyn Evans} \& {Gilmore}}{{Walker} et~al.}{2009}]{Walker09}
{Walker} M.~G.,  {Mateo} M.,  {Olszewski} E.~W.,  {Pe{\~n}arrubia} J.,  {Wyn
  Evans} N.,    {Gilmore} G.,  2009, \apj, 704, 1274

\bibitem[\protect\citeauthoryear{{Walker}, {Mateo}, {Olszewski},
  {Pe{\~n}arrubia}, {Wyn Evans} \& {Gilmore}}{{Walker} et~al.}{2010}]{Walker10}
{Walker} M.~G.,  {Mateo} M.,  {Olszewski} E.~W.,  {Pe{\~n}arrubia} J.,  {Wyn
  Evans} N.,    {Gilmore} G.,  2010, \apj, 710, 886

\bibitem[\protect\citeauthoryear{{Wang} \& {White}}{{Wang} \&
  {White}}{2007}]{ww07}
{Wang} J.,  {White} S.~D.~M.,  2007, \mnras, 380, 93

\bibitem[\protect\citeauthoryear{{Wolf}, {Martinez}, {Bullock}, {Kaplinghat},
  {Geha}, {Mu{\~n}oz}, {Simon} \& {Avedo}}{{Wolf} et~al.}{2010}]{wo10}
{Wolf} J.,  {Martinez} G.~D.,  {Bullock} J.~S.,  {Kaplinghat} M.,  {Geha} M.,
  {Mu{\~n}oz} R.~R.,  {Simon} J.~D.,    {Avedo} F.~F.,  2010, \mnras, 406, 1220

\bibitem[\protect\citeauthoryear{{Xue}}{{Xue} et~al.}{2008}]{Xue_08}
{Xue} X.~X., et al.,  2008, \apj, 684, 1143


\end{thebibliography}

\bsp

\label{lastpage}

\end{document}